# Notes to Saturn satellites Ijiraq and Kiviuq mutual close encounters

A.E. Rosaev

FGUP NPC "NEDRA", Yaroslavl, Russia

The problem of origin of outer irregular satellites of large planets is considered. The capture way of their origin most probable, however there is not detail theory. There are a number of irregular satellites, discovered in recent time. It gives an ability to investigate the statistics of orbital interaction and try to reconstruct real collision history of these objects We restrict this consideration by pair of orbits with close elements: Kiviuq and Ijiraq and determine period of close encounters between this satellites.

## 1. INTRODUCTION

Nesvorny et all [1] research evolution of orbit orientations in asteroid families and prove fact of recent catastrophic destruction in asteroid main belt. On the other hand, the method of investigations of genetic relations between minor bodies successfully applied in our previous works [2]. It is naturally to apply these two ideas to irregular satellites orbits with close elements for determination their parent body possible catastrophic destruction epoch.

The problem of determination of the minimal distance between elliptic orbits reduced to solving the system of equations [2]:

$$\Delta^2 = (x_1 - x_0)^2 + (y_1 - y_0)^2 + (z_1 - z_0)^2$$

$$x_j = R_j (\cos u_j \cos \Omega_j - \sin u_j \sin \Omega_j \cos i_j)$$
$$y_j = R_j (\cos u_j \sin \Omega_j + \sin u_j \cos \Omega_j \cos i_j)$$
$$z_j = R_j \sin u_j \sin i_j$$

$$u_j = \lambda_j - \Omega_j, \quad j = 0,1$$

Last expression is possible to modify as follows. Order with the average distance between orbits important value has a value of the minimum distance.

We suppose the new method of numeric search of this system, based on a suggestion, that minimal distance reach close to point, determined by conditions:

$$\lambda_1 = \vartheta_1 - \omega_1 - \Omega_1 = \lambda_2 = \vartheta_2 - \omega_2 - \Omega_2 = \lambda \qquad (1)$$
$$r_1 = r_2 = r$$

More, the expressions for minimal distance between orbits determination:

$$\Delta_s^2 = r^2 \sum_{i=1}^{n} (a_i \cos\lambda + b_i \sin\lambda)^2 \qquad (2)$$

where:

$$\lambda_{min} = 0{,}5 \operatorname{atg} \frac{2\sum a_i b_i}{\sum a_i^2 - b_i^2} \qquad (3)$$

$$a_1 = \cos^2\Omega_1 + \sin^2\Omega_1 \cos i_1 - \cos^2\Omega_2 - \sin^2\Omega_2 \cos i_2$$
$$b_2 = \sin^2\Omega_1 + \cos^2\Omega_1 \cos i_1 - \sin^2\Omega_2 - \cos^2\Omega_2 \cos i_2$$
$$a_2 = \sin\Omega_1 \cos\Omega_1 (1 - \cos i_1) - \sin\Omega_2 \cos\Omega_2 (1 - \cos i_2)$$
$$b_1 = \sin\Omega_1 \cos\Omega_1 (1 - \cos i_1) - \sin\Omega_2 \cos\Omega_2 (1 - \cos i_2)$$
$$a_3 = -\sin\Omega_1 \sin i_1 - \sin\Omega_2 \sin i_2$$
$$b_3 = \cos\Omega_1 \sin i_1 - \cos\Omega_2 \sin i_2$$

It was found 8 roots for the determination of the minimum distance between elliptical orbits and given way of their calculation. So naturally place in (2) n=8.

It allows us to enter a new criterion for genetic related orbits search:

$$DN = \sum \Delta_i / n$$

where $\Delta_i$ determined (2)-(3). The similar criterion was supposed by Danielsson [3], but presented consideration have some advantages. First of all - it clear dependence from time.

The quasi coorbital objects have a particular interest (tables1, 2). It is natural to suppose, that such close orbits appear not causal, but a product of catastrophic breakup with small relative velocity. For such objects a mean distance between orbits is introduced, where points of orbital distance calculation chosen by special way.

As a result, the possible epochs of few most recent catastrophic breakups may be determined. The satellites elements of orbits, according to these catastrophic events, may be estimated. It is evident, that breakup process may be related with capture; and this process could be multistage.

There are few groups of irregular satellites in Jupiter and Saturn systems. We restrict this consideration by pair of orbits with close elements (table 1 – 2, Fig 1).

Table 1

Saturn satellites with close orbits mean elements (R. A. Jacobson, private communications)

|   |        | n (deg/day) | a         | e      | w      | i      | node    | size |
|---|--------|-------------|-----------|--------|--------|--------|---------|------|
| 1 | Kiviuq | 0.8013**901** | 1.1365000 | 0.3336 | 83.043 | 46.148 | 359.519 |      |
| 2 | Ijiraq | 0.7974**023** | 1.1442000 | 0.3215 | 85.523 | 46.73  | 136.348 |      |

Table 2

Main periods at Jovian and Saturnian satellites with close orbits

|          | P days                | $T_{con}$, yr | Epoch of the most recent conjunct | **Pnode, yr** | $T_{node}$, yr | Epoch of orbit optimal orientation for intersections | ΔT between two nearest conjunctions at epoch of intersections |
|----------|-----------------------|---------------|-----------------------------------|---------------|----------------|-------------------------------------------------------|----------------------------------------------------------------|
| Elara    | 259.64                | 418.0         |                                   | 252.49        | 1345.07        |                                                       | 562238.66                                                      |
| Lysithea | 259.20                |               |                                   | 299.88        |                |                                                       |                                                                |
|          |                       |               |                                   |               |                |                                                       |                                                                |
| Kiviuq   | 449.21942 ±0.00006    | 245.7±0.2     | 11.532                            | 846.46        | 32880±30       | 12879.05                                              | $(8.08±0.01)*10^6$                                             |
| Ijiraq   | 451.46596 ±0.00006    |               |                                   | **868.25**    |                |                                                       |                                                                |

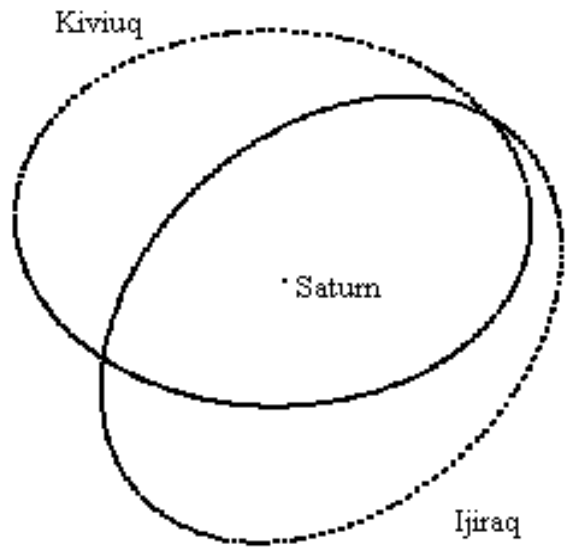

Fig.1 Ijiraq And Kiviuq orbit projection at present

## 2. EPOCH OF IJIRAQ AND KIVIUQ PARENT BODY DESTRUCTION DETERMINATION

The studying outer natural satellite close orbits rather easy than asteroids by some reasons first of all, the main perturbation from the Sun change in more narrow range, than Jupiter perturbation for minor planets, and there are no subsequent encounters with another objects.

The orbital periods and rates of orbit precession are given in table 2. The period Tcon between conjunction and period Tnode between two similar orbit orientations are calculated by expressions:

$$Tcon = T1/(T2-T1)*T1/365.24$$

$$Tnode = T1/(T2-T1)*T1$$

and it was verified by numeric calculations.

It is evident, that for breakup epoch the distance between orbits must be close to zero, and conjunction (around longitude of minimal distance) must take place. Let breakup take place at epoch $t$. Next ability to exact repeat breakup configuration will take place through about 33000 years due to mutual node motion (orbit precession), however, due to long time between subsequent conjunctions (240 years), not each configurations is suitable. The variation of difference between epoch of orbit exact intersection and nearest conjunction Δ is given at Fig.2 and table 3. A very long period between conjunction and orbit configuration repeating for Kiviuq and Ijiraq case (deltaT) means, that we have relative small number of similar events (about 500) at the history of the Solar system and can to consider each more carefully. Due to Yarkovsky effect, close orbits can change their mutual positions, so most recent cases of close encounters are most possible for breakup.

Moreover, if conjunction takes place close to epoch 33000 year, according to precession period, it can take place far from longitude of orbit intersection point. It means, that the longitude of minimal distance between orbits must take into account in ΔT value (table 2). The calculations show that really zero encounters between Saturnian moonlets can take place about one time per 100 million years. So, we must consider only about 40 cases!

Denote:

$$a = \sin(2\pi / T_{con} \, t - E_{0\,con})$$

$$b = \sin(2\pi / T_{node} \, t - E_{0\,node})$$

At the moment of breakup $a= 0$ and $b= 0$.

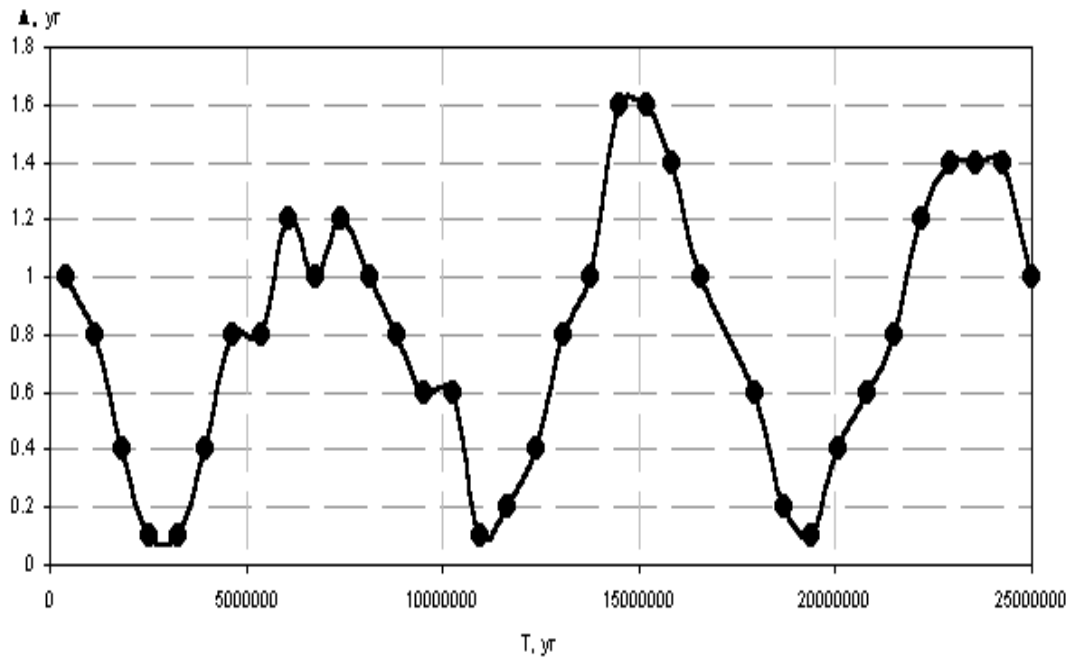

Fig.2 The variation of difference between epoch of orbit exact intersection and nearest conjunction

Table 3

The variation of difference between epoch of orbit exact intersection and nearest conjunction

| Epoch, yr | $\Delta$, yr |
|---|---|
| 2540938.6 | 0.2 |
| 3246825.2 | 0.1 |
| 10962331.2 | 0.1 |
| 19383723.6 | 0.025 |
| 27805116.3 | 0.05 |
| 35520622.3 | 0.15 |
| 36226508.8 | 0.20 |

In according with table 3 data, main result of this investigation – a most possible Epoch of catastrophic breakup when Kiviuq and Ijiraq appear is $1.938 \pm 0.002 \times 10^6$ yr. The variation of initial orbital elements and precession rate was applied.

By what way the precision of may be improved? 1) We must to consider perturbation in inclination. 2) a more carefully consideration of longitudes conjunctions and longitude conjunction

CONCLUSIONS

The unique orbital characteristics of small irregular Saturn satellites Ijiraq and Kiviuq provide a very rare mutual close encounters . It is possible to calculate main periods in studied system. The epoch of most recent close encounter is calculated.

References:

1. Rosaev A.E., 2001. **The reconstruction of genetic relations between minor planets based on their orbital characteristics,** *In: Dynamics of Natural and Artifical Celestial bodies,- Proceedings of US-European Celestial Mechanics Workshop, 3-7 July 2000, Poznan, - ed. Pretka-Ziontek et al., Kluwer Acad. Pub., p. 301-303.*
2. Nesvorny D., Bottke W. F. Jr, Dones L. & Levison H. F., 2002. **The recent breakup of an asteroid in the main-belt region.** *Nature, Vol 417, 13 June 2002, P.721*
3. Danielsson L. Statistical arguments for asteroidal jet streams. - *//Astroph. Space Sci..- 5.- P.53-58*